\documentclass{SciPost}

\hypersetup{
    colorlinks,
    linkcolor={red!50!black},
    citecolor={blue!50!black},
    urlcolor={blue!80!black}
}

\newcommand{\changemore}[1]{{\color{black}#1}} %{{#1}}

\usepackage[bitstream-charter]{mathdesign}
\usepackage{subfig}
\usepackage[normalem]{ulem}
\DeclareSymbolFont{usualmathcal}{OMS}{cmsy}{m}{n}
\DeclareSymbolFontAlphabet{\mathcal}{usualmathcal}

\fancypagestyle{SPstyle}{
\fancyhf{}
\lhead{\colorbox{scipostblue}{\bf \color{white} ~SciPost Physics }}
\rhead{{\bf \color{scipostdeepblue} ~Submission }}

\fancyfoot[C]{\textbf{\thepage}}
}

\begin{document}

\pagestyle{SPstyle}

\begin{center}{\Large \textbf{\color{scipostdeepblue}{
%%%%%%%%%% TODO: Write your article's title here
Many-body localization in a quantum Ising model with the long-range interaction: Accurate determination of the transition point\\
}}}\end{center}

\begin{center}\textbf{
% Use (full) first name (+ middle name initials) + surname format.
% Separate subsequent authors by a comma, omit comma and use "and" for the last author.
% Mark the corresponding author(s) with a superscript symbol in this order
% \star, \dagger, \ddagger, \circ, \S, \P, \parallel, ...
Illia Lukin\textsuperscript{1},
Andrii Sotnikov\textsuperscript{1,2} and
Alexander L.  Burin\textsuperscript{3$\star$}
}\end{center}

\begin{center}
{\bf 1} National Science Center "Kharkiv Institute of Physics and Technology", Akademichna str. 1, 61108 Kharkiv, Ukraine
\\
{\bf 2} V.N. Karazin Kharkiv National University,  Svobody Square 4, 61022 Kharkiv, Ukraine
\\
{\bf 3} Department of Chemistry, Tulane University, New
Orleans, LA 70118, USA 
%%%%%%%%%% END TODO: AFFILIATIONS
% Provide email address of corresponding author(s)
\\[\baselineskip]
$\star$ \href{mailto:aburin@tulane.edu}{\small aburin@tulane.edu}\
%$\star$ \href{mailto:email1}{\small email1}\,,\quad
%$\dagger$ \href{mailto:email2}{\small email2}
\end{center}

\section*{\color{scipostdeepblue}{Abstract}}
\textbf{\boldmath{%
We investigate the many-body localization (MBL) transition in the quantum Ising model with long-range interactions. Unlike spin chains with short-range interactions, where the MBL transition point remains elusive due to strong finite-size effects and local fluctuations, long-range interactions suppress such fluctuations and enable clearer signatures of critical behavior. Using exact results from a related Bethe lattice localization problem, we estimate the MBL threshold within logarithmic accuracy and find consistency with exact diagonalization. Although the critical disorder diverges in the thermodynamic limit, our results demonstrate that the critical regime can still be probed and highlight the relevance of this model to systems with dipole–dipole, elastic, or indirect exchange interactions. 
}}

\vspace{\baselineskip}

%%%%%%%%%% BLOCK: Copyright information
% This block will be filled during the proof stage, and finilized just before publication.
% It exists here only as a placeholder, and should not be modified by authors.
\noindent\textcolor{white!90!black}{%
\fbox{\parbox{0.975\linewidth}{%
\textcolor{white!40!black}{\begin{tabular}{lr}%
  \begin{minipage}{0.6\textwidth}%
    {\small Copyright attribution to authors. \newline
    This work is a submission to SciPost Physics. \newline
    License information to appear upon publication. \newline
    Publication information to appear upon publication.}
  \end{minipage} & \begin{minipage}{0.4\textwidth}
    {\small Received Date \newline Accepted Date \newline Published Date}%
  \end{minipage}
\end{tabular}}
}}
}
%%%%%%%%%% BLOCK: Copyright information

% For convenience during refereeing we turn on line numbers:
%\linenumbers

\vspace{10pt}
\noindent\rule{\textwidth}{1pt}
\tableofcontents
\noindent\rule{\textwidth}{1pt}
\vspace{10pt}

% Write your article contents here, starting from first \section.
% An example structure is given below.

\section{Introduction}
\label{sec:intro}
% TODO: write your article here.

In contrast to single-particle localization~\cite{Anderson58}, which emerges when interactions are negligible and typically occurs at zero temperature, many-body localization (MBL) can arise at arbitrary temperatures. The MBL transition separates chaotic (delocalized) and deterministic (localized) dynamical phases in interacting quantum systems. In the chaotic phase, the eigenstate thermalization hypothesis (ETH) holds \cite{Srednicki94ETH,Deutsch91},  meaning the system acts as a thermal bath for its subsystems and reaches thermal equilibrium internally. By contrast, the localized phase fails to thermalize, and the system retains memory of its initial state indefinitely. In the chaotic regime, equilibration involves energy equipartition and irreversible relaxation that erases the system’s initial information, whereas the localized phase preserves it. This robustness of memory makes the localized phase particularly desirable for quantum technologies, including quantum hardware  \cite{Debnath2016,Britton2012,Parameswaran2018Review,Altman2018Qinf,
Monroe2021observationStarkMBL,Polkovnikov11RevModPhys}, where relaxation processes are a fundamental source of decoherence.

Despite extensive efforts to understand and characterize the many-body localization (MBL) transition in various systems~\cite{FleishmanAnderson80, KaganMaksimov84, MIKHEEV83He3He4, KaganMaksimov85, ab90Kontor,
Wolynes2, LeitnerArnoldDiffusion97, Shepelyansky98, Gornyi05,
Basko06, OganesyanHuse07, Imbrie16Review} two fundamental questions remain unresolved: (A) How can the MBL transition point be determined? and (B) What is the nature of the system’s behavior in the critical regime near the transition? For single-particle localization, the transition threshold can be reliably identified numerically since the complexity of the eigenstate basis grows only polynomially with system size \cite{EversMirlin08}.  In contrast, for MBL the complexity grows exponentially \cite{Wolynes2,AltshullerGefen97}, preventing convergence in numerically accessible spin chains with short-range interactions, even though the existence of the transition in the thermodynamic limit has been rigorously established \cite{Imbrie16,Imbrie16Review}. Here, we investigate the MBL transition in spin systems with long-range interactions, where the transition can be analyzed analytically by exploiting its similarity to the solvable localization problem on the Bethe lattice \cite{AbouChacra73}. This approach provides a well-defined transition point and opens the way to probing critical behavior in the vicinity of the transition.

Generally, MBL problem can be formulated using the Hamiltonian split into static and dynamic parts as \cite{KaganMaksimov85,ab90Kontor} 
\begin{eqnarray}
\widehat{H}=\widehat{H}_{0}+\xi \widehat{V}
\label{eq:HComp}
\end{eqnarray}
where $\widehat{H}_{0}$ is a static Hamiltonian  composed of commuting operators $\widehat{A}_{i}$ (e.g., spin projection operators   $S_i^{z}$ to the $z$ axis defined in local sites $i$) and $\widehat{V}$ is a dynamic interaction  including operators not commuting with each other and the static Hamiltonian (e.g., spin projection operators~$S_i^{x}$ to the $x$ axis, as in Eq.~(\ref{eq:Hrl}) below). The parameter~$\xi$ (transverse field~$\Gamma$ in Eq.~(\ref{eq:Hrl})) controls the strength of dynamic interaction.  The system is obviously localized for $\xi=0$, and the physical parameters characterized by the operators $\widehat{A}_{i}$ serve as its integrals of motion. The static eigenstate basis $\{B\}$ of the system Fock space is chosen using the states with given values of operators $\widehat{A}_{i}$ (for $N$ interacting spin-$1/2$ particles this will be $2^{N}$ states, characterized by different spin projections $S_{i}^{z}=\pm 1/2$ to the $z$ axis).  The Fock space for a single-particle localization problem involves only $N$ states defined, for example, as a single spin excitation of the ferromagnetic state with the total spin projection to the $z$ axis $\pm (N/2-1)$, provided that the dynamic interaction~$\widehat{V}$ does not modify the spin projection.  

When the dynamical parameter $\xi \neq 0$, dynamical interactions modify the eigenstates, which become superpositions of static-basis states with fixed operators $A_{i}$. In the single-particle problem, this modification can be quantified by the number of static-basis states contributing to a single eigenstate: if this number remains independent of system size, the states are localized, whereas if it grows with system size, the states are delocalized. The chaotic (delocalized) regime is further characterized by Wigner-Dyson level statistics~\cite{ShklovskiiShapiro93},  reflecting level repulsion. In contrast, localized states are spatially separated and exhibit Poisson level statistics, as level repulsion is absent.  At the transition point, eigenstates display marginal behavior between these two regimes.

It has been rigorously established~\cite{Imbrie16} that the MBL phase can emerge in certain spin chains with short-range interactions at sufficiently small but nonzero dynamical interactions ($\xi \neq 0$). However, advanced numerical studies of the localization threshold in such chains, even for system sizes up to $N = 25$ spins \cite{Sierant2022Challenges, Sierant2020LargeMatr,2024SierantMBLShortRange}, show very slow convergence, likely due to sensitivity to local fluctuations~\cite{Huveneers17BrekDwnLoc}, which can drive localization breakdown. Conclusive results would require simulations of systems larger than $30$ spins, beyond the reach of classical computation. Moreover, no analytical method exists to determine the localization threshold in MBL with short-range interactions, paralleling the challenge of the Anderson localization problem.

\changemore{We expect the MBL problem to be more tractable in systems with long-range interactions. This expectation is supported by numerical studies of single-particle Anderson localization in high spatial dimensions 
$d\geq 3$ \cite{GarciaCuevas07} with the large number of neighbors and on the Bethe lattice~\cite{AbouChacra73} with the large coordination number, where the localization transition occurs when there is, approximately (up to logarithmic accuracy), one resonant coupling between eigenstates of the quasi-static Hamiltonian $H_0$  induced by the perturbation $\xi V$, Eq.~(\ref{eq:HComp}). Furthermore, the localization threshold on the Bethe lattice is known analytically \cite{AbouChacra73}, and is consistent with the advanced numerical investigations \cite{sierant2023RRG,2023TikhonovRRG, 2024RGRRG}.   

In the present work, we investigate the MBL problem in a system of $N$ interacting spins coupled through random, static, binary interactions that do not depend on the distance between spins~\cite{ab16SG}. We argue that the localization threshold in this model is likewise determined, up to a logarithmic factor, by the condition of having approximately one resonance per many-body state. Using level-statistics analysis  \cite{OganesyanHuse07}, we determine the critical disorder and find excellent agreement with this analytical estimate. Following earlier studies suggesting the approximate relevance of Bethe-lattice physics to the MBL problem \cite{LoganWolynes90,AltshullerGefen97}, we construct a matching Bethe-lattice model and show that the localization threshold in our system is roughly a factor of three smaller than in the corresponding Bethe-lattice problem. Establishing this threshold provides a solid foundation for future investigations of critical behavior near the MBL transition.}

One additional comment is in order.  The long-range interaction is almost ubiquitous in any system of interacting quantum objects  (spins, electrons or atomic tunneling systems) due to dipole-dipole, elastic, magnetic dipole or indirect exchange interactions \cite{ab98book,ab06preprint,ab15MBL}.  Therefore, the present work is potentially applicable to many physical systems, where many-body localization has been observed, including interacting two-level systems in amorphous solids \cite{ab95TLSRelax,ab98book,ab98prl},  nitrogen vacancies in diamond~\cite{LukinDiamond16}, and cold ions~\cite{Monro16}. Certainly, these systems were analyzed theoretically and reported in numerous publications \cite{ab90Kontor,ab98book,ab06preprint,Lukin14MBLGen,ab15MBL,ab16GutmanMirlin,abGorniyMirlinDot,ab16SG,Nandkishore17lr,GargSK18}, and there exists common understanding of the nature of MBL transition based on the criterion of a single resonance per state similar to that for the Anderson localization problem in three dimensions. Within the present study, we attempt to characterize MBL transition not only at the qualitative, but quantitative level estimating accurately the MBL threshold.

\changemore{The paper is organized as follows. In Sec.~\ref{sec:Model}, we introduce the interacting-spin model and the corresponding matching Bethe-lattice problem. In Sec.~\ref{sec:analitTh}, we outline the analytical expectations for the MBL transition at infinite temperature, following earlier work by one of the authors \cite{ab16SG}. Section~\ref{sec:ExactNum}  presents a numerical study of the localization problem based on level-statistics analysis and compares the results with the analytical predictions for the matching Bethe-lattice model. Finally, Sec.~\ref{sec:Concl} summarizes the main findings of the present work.}

\section{Model and Bethe lattice counterpart}
\label{sec:Model}

\begin{figure}[t]
\centering
 \includegraphics{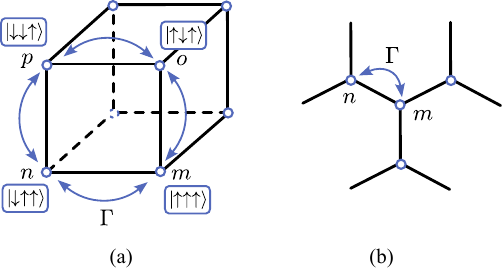}
%\subfloat[]{\includegraphics[scale=0.3]{Fig1a.eps}}%[scale=0.4]{Fig1a_upd1.pdf}%
 %\hspace{1cm}
%\subfloat[]{\includegraphics[scale=0.4]{Fig1b.eps}}%[scale=0.4]{Fig1b_upd1.pdf}%
{\caption{\small (a) Hypecube with vertices representing the phase space of three spins.  (b) Bethe lattice with each site coupled to three neighbors.}
\label{fig:HCBL}}
\end{figure}

We investigate the quantum spin glass model of $N$ interacting spins characterized by the  Hamiltonian   having a form similar to  Eq. (\ref{eq:HComp}) (see Refs. \cite{Laumann14,ab16SG}) \changemore{
\begin{eqnarray}
{H}=H_0+V, \quad H_0=\frac{1}{2}\sum_{i\neq j}J_{ij}S_{i}^zS_{j}^{z}, \quad V=2\Gamma \sum_{i}S_{i}^x,
\label{eq:Hrl}
\end{eqnarray}}
where $S_{i}^{\mu}$ are spin $1/2$ operators ($\mu=x$, $y$, or $z$). The first term \changemore{$H_0$} is the static Hamiltonian representing the celebrated Sherrington-Kirkpatrick model \cite{Sherrington75} with random, uncorrelated interactions $J_{ij}$ characterized by the Gaussian distribution with the mean $\langle J_{ij}\rangle=0$ and $\langle J_{ij}J_{kl}\rangle=J_{0}^2\left(\delta_{ik}\delta_{jl}+\delta_{il}\delta_{jk}\right)$, while the second term \changemore{$V$} corresponds to the transverse field.  In the absence of the dynamic term ($\Gamma=0$), eigenstates of the model are represented by the $2^{N}$ vertices of $N$-dimensional hypercube with each vertex coordinates given by $N$ spin projections $S^{z}=\pm 1/2$ for the corresponding state, as illustrated in Fig.~\ref{fig:HCBL}(a) for three spins. Previously, the MBL problem has been investigated for a random energy model \cite{Derrida1981,Laumann14}. The phase space of this model can be similarly represented by a hypercube. Yet, the quantum random energy model is equivalent to the Anderson localization problem on the hypercube, while the present model possesses a strong correlation between different state energies \cite{ab16SG} and destructive interference, similarly to the MBL problems with short-range interactions. We consider the present model as a bridge between single-particle and many-body localization problems.

\changemore{We do not normalize the spin-spin interaction by the factor $1/\sqrt{N}$ as in the original Sher\-ring\-ton-Kirkpatrick model \cite{Sherrington75}, because this normalization was necessary to keep the spin glass transition independent of the number of spins $N$, while for the present problem it still leads to the localization threshold approaching zero. }

 \changemore{The energy difference between adjacent hypercube vertices $m$ and $n$ (see Fig.~\ref{fig:HCBL}(a))  is given by a spin-flip energy
\begin{eqnarray}
\omega_{i}=2S_i^z\sum_{j}J_{ij}S_j^z.    
\end{eqnarray}
We are interested in the MBL transition at an infinite temperature for the states with total energy close to zero.  Then the correlations between different spin-flip energies can be approximately neglected for the large number of spins since the correlation function of different spin flip energies $\langle\omega_{i}\omega_{j}\rangle=J_{ij}^2/4 \sim J_0^2/4$ is much smaller than the average squared energy $\langle\omega_{i}^2\rangle=\sum_{j}J_{ij}^2/4\approx NJ_0^2/4$.  Consequently, the site energies are correlated, while their differences are approximately uncorrelated. The distribution of spin-flip energies is given by  the  Gaussian distribution ($N\gg 1$) 
\begin{eqnarray}
P(\omega)=\frac{1}{\sqrt{2\pi}W}e^{-\frac{\omega^2}{2W^2}}, \quad W=\sqrt{N-1}\frac{J_{0}}{2}. 
\label{eq:randp}
\end{eqnarray}}

 %(Gaussian distribution with the width defined as $W=\sqrt{N-1}J_{0}/2$).  

The matching Bethe lattice problem is constructed following the recipe of Refs.~\cite{Wolynes2,ab16SG,Logan2018}.  Each state of the spin system forms the vertex in the graph coupled with $N$ other states different from this by a single spin flip and the coupling strength is given by the dynamic parameter~$\Gamma$.  Thus, the number of neighbors in the matching Bethe lattice model is $N$.  \changemore{Similarly to the problem of interest, Eq.~(\ref{eq:Hrl}), we assume that the energies of different sites are correlated while their differences are not.  Then, as argued in Ref.~\cite{ab16SG}, the random potential probability density function  $g(E)$ entering the localization threshold for the states with the energy $E$ on the Bethe lattice with uncorrelated site energies is replaced with the probability $P(0)$, Eq.~(\ref{eq:randp}), that the neighboring site is in resonance with the given site. The more accurate justification of that is given in Ref. \cite[Supplementary Materials]{ab17SGpreprintSI}.} %, Supporting Information

\changemore{Therefore, the localization threshold in the Bethe lattice problem described above in the limit $N\gg 1$ is determined as \cite{AbouChacra73} 
\begin{eqnarray}
    1 &\approx& 4(N-1)\frac{\Gamma_{\rm B}}{\sqrt{2\pi}W}\ln\left(\frac{\eta_{\rm B} W}{\Gamma_{\rm B}}\right),
    \label{eq:BLThr}
\end{eqnarray}
where $\eta_{\rm B}$ is an unknown parameter of the order of unity. We use $\eta_{\rm B} \approx 1.06$, as  estimated in Appendix (see Eq.~(\ref{eq:AppBLThr})) following Ref. \cite{AbouChacra73} for the Bethe lattice problem with the Gaussian disorder, which should be approximately relevant for the present situation.} 

Equation~(\ref{eq:BLThr}) determines the localization threshold of the Bethe lattice problem, and we use it below to evaluate the localization threshold for the matching MBL problem in the interacting spin system characterized by the Hamiltonian~(\ref{eq:Hrl}). 

\section{Analytical estimate of the MBL threshold}
\label{sec:analitTh}

The phase space is organized differently for the quantum Ising model shown in Fig.~\ref{fig:HCBL}(a) and the corresponding  Bethe lattice problem shown Fig.~\ref{fig:HCBL}(b). In the former case, the graph representing the system (hypercube)  and it has plenty of loops, while there are no loops in the latter case.  Loops result in destructive interference for the hypercube, which is absent in the  Bethe lattice~\cite{ab16SG}.   

The destructive interference is realized already in the second-order process involving two spins flips illustrated  in Fig.~\ref{fig:HCBL}(a) as the transition between the initial state $m$ and the final state $p$ that can happen by means of the flip of spins $i$ and $j$.   If the spin $i$ transfers first, then the transition goes through the vertex $n$ as $m \rightarrow n \rightarrow p$, while in the opposite case it goes through the vertex $o$ as $m \rightarrow o \rightarrow p$.  The total transition amplitude in the second order of perturbation theory in the dynamic field $\Gamma$ is given by the sum of contributions of two paths, which is defined as $\Gamma_{ij}=\Gamma^2(\omega_{i}^{-1}+\omega_{j}^{-1})$ (remember that notation $\omega_{k}$ refers to  the flip energy of the spin $k$).  

The second-order process is significant in the resonant regime, where the total energy change due to two spin flips $\omega_{ij}=\omega_{i}+\omega_{j}-J_{ij}^{*}$ ($J_{ij}^{*}=4J_{ij}S_{i}^{z}S_{j}^{z}$ and spin projections $S_{i}^{z}$, $S_{j}^{z}$ corresponds to the static basis state $m$) approaches zero.  In this resonant regime, the second-order transition amplitude reads 
\begin{eqnarray}
    \Gamma_{ij}= \frac{\Gamma^2J_{ij}^{*}}{\omega_{i}\omega_{j}}.
    \label{eq:SecondOrdProc}
\end{eqnarray}
The second-order transition amplitude approaches zero in the absence of interaction between spins $i$ and $j$, reflecting the many-body nature of the two spin flip contribution lacking if the spins are just in static longitudinal fields, where no MBL transition can, obviously, take place.  

The second-order amplitudes  are substantially suppressed, if the spin-flip energies are large compared to the interaction between spins, i.e., $|\omega_{i}| \approx |\omega_{j}| \gg |J_{ij}^{*}|$. In the opposite cases of either $|\omega_{i}| \ll |J_{ij}^{*}|$ and $|\omega_{j}| \approx  |J_{ij}^{*}|$ or $|\omega_{j}| \ll |J_{ij}^{*}|$ and $|\omega_{j}| \approx  |J_{ij}^{*}|$, the contributions of two paths do not interfere with each other and contribute as two independent paths similarly to that in the Bethe lattice.  This suggests the constraint of the maximum spin-flip energy in the definition of the localization threshold given by Eq.~(\ref{eq:BLThr}) by the spin-spin interaction $J_{0}/2$.  Setting this constraint in the integral in Eq.~(\ref{eq:BLThr}) we obtain the equation for localization threshold within the logarithmic accuracy in the form 
\begin{eqnarray}
    \beta\approx 4(N-1)\frac{\Gamma_{c}}{\sqrt{2\pi}W}\ln\left(\frac{\eta J_{0}}{\Gamma_{c}}\right),  
    \label{eq:MBLThr}
\end{eqnarray}
where $\beta$ and $\eta$ are unknown parameters of the order of unity and it is expected that $\beta>1$~\cite{ab16SG}. The parameter $\beta >1$ originates from the higher-order processes determining the localization threshold. It is equal to unity for the localization problem on the Bethe lattice, but here it can be different because of the difference in the total number of distinguishable system pathways.  Indeed, the total number of $n$-step paths  in the Bethe lattice from the given point scales as $N(N-1)^{n-1}$, while for the interacting spin problem it scales as $N!/(N-n)!$ for the number of pathways~\cite{ab16SG}. Consequently, if the MBL transition is determined by the higher-order processes of the order of $N$, then the critical dynamic field $\Gamma_{c}$ can increase due to the reduction in the number of pathways, as reflected by the additional parameter $\beta>1$ in Eq.~(\ref{eq:MBLThr}). 

Our derivation assumes that $\Gamma < J_{0}$, which is obviously true near the MBL threshold, since the width~$W$ scales with the number of spins as $J_{0}\sqrt{N}$. Thus, we get $\Gamma_{c}^{\rm MBL} \sim J_{0}/(\sqrt{N}\ln(N))$ similarly to the Bethe lattice.  Consequently, $\Gamma_{c} \ll J_{0}$ in the regime of interest $N \gg 1$.  

\changemore{We expect that consideration of higher-order processes does not modify the analytical form definition of the localization threshold in Eq. (\ref{eq:MBLThr}), but only modifies  the numerical parameters $\beta$ and $\eta$.  As it was argued in the earlier work of one of us \cite{ab16SG} (see also the analysis in greater detail in Ref.~\cite[Supplementary Materials]{ab17SGpreprintSI}), the consideration of higher ($n^{th}$) order spin transitions within the present model rescales  the argument of logarithm in the definition of the localization threshold by the square root of the number of spin flips $n$. Such changes together with the reduction in the number of non-self intersecting paths compared to the number of such paths in the Bethe lattice are expected to modify only numerical factors compared to that for the Bethe lattice. These expectations are confirmed by the numerical results reported below in Sec. \ref{sec:ExactNum}. 

In case of short-range spin-spin interactions, most of the second-order contributions given by Eq.~(\ref{eq:SecondOrdProc}) vanish, because the associated spin-spin interaction $J_{ij}$ is equal zero for them. Consequently, the present consideration is not applicable to the models with short-range interactions, where processes of high order in the number of spin flips are dramatically significant for understanding the MBL transition \cite{Sierant2022Challenges,villalonga2020eigenstateshybridizelengthscales,2021FloqShrMBL,2022MBLThCross,2023POhenPrethLong}.} 

%Below we report the investigation of MBL transition  using the exact diagonalization of the Hamiltonian~(\ref{eq:Hrl}).  

\section{ Exact diagonalization results for the MBL threshold}
\label{sec:ExactNum}

\changemore{In typical MBL problems, the number of static-basis states contributing to a single eigenstate increases with system size even in the localized regime~\cite{Serbyn19Review}, though at a slower rate than the growth of the full basis set. The onset of Wigner–Dyson statistics signals the chaotic regime~\cite{OganesyanHuse07,Serbyn19Review}, where all states effectively couple and strong level repulsion emerges. Alternative criteria for identifying the MBL transition include the long-time behavior of local correlation functions~\cite{BerezinskiiGorkov79,ab15MBL}, the presence of local integrals of motion \cite{Serbyn13LocalIntMot}, and the time evolution of entanglement entropy~\cite{Pollman12EntEntr}.  Among these, level-statistics analysis is the most computationally accessible, as it requires only the eigenenergies of selected states, which can be obtained through sparse-matrix diagonalization methods applicable to relatively large systems. Therefore, our consideration below is based on the level statistics.}

To determine quantitatively the critical field strength $\Gamma_{c}$, we analyze the level statistics expressed through the gap ratio parameter $ r$ \cite{OganesyanHuse07}
\begin{eqnarray}
     r  =  \frac{\min\left(\Delta_n,\Delta_{n+1}\right)}{\max\left(\Delta_n,\Delta_{n+1}\right)} ,
    \label{eq:lst} 
\end{eqnarray}
where the quantities $\Delta_n=E_{n+1}-E_{n}$ represent the differences in energies $E_{n}$ of the adjacent eigenstates (energy gaps). This ratio is then averaged over some energy window.  
The localized phase is characterized by the average gap ratio $ r  \approx 0.3863$, while in the delocalized (chaotic) phase $ r  \approx 0.5307$, where the average of the gap ratio is computed either over the full energy spectrum or over the specified energy window for a fixed choice of disorder. 

\begin{figure}[t]
    \centering
    \includegraphics[width=0.6\linewidth]{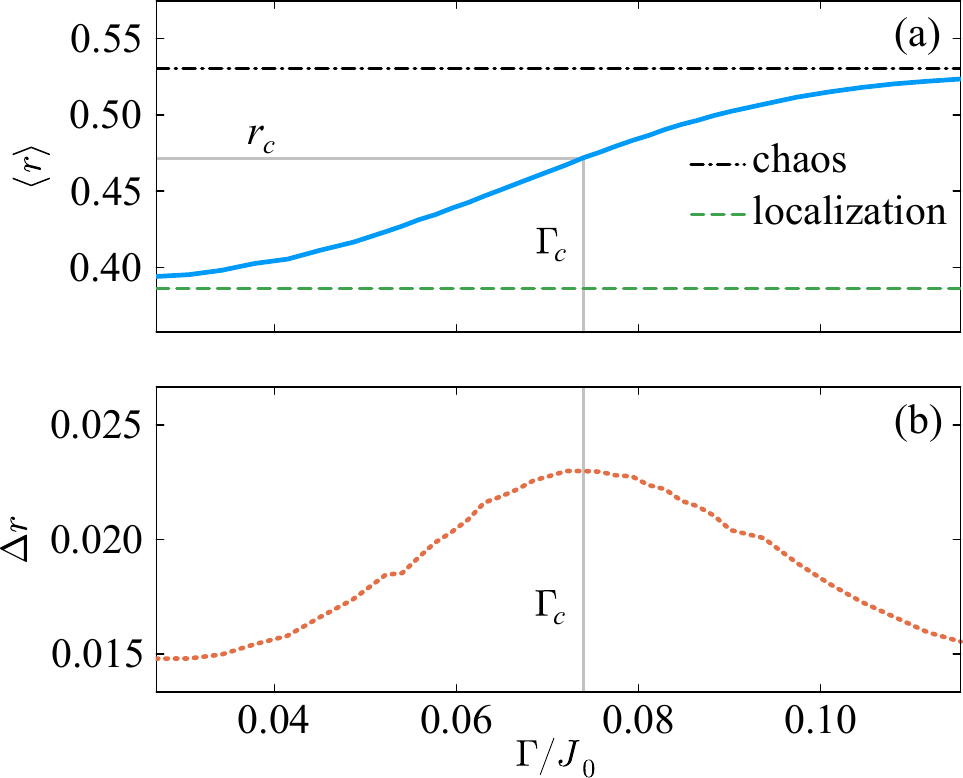}      \caption{\label{fig:Illustration}%
     General behavior of \changemore{(a) the average gap ratio} $\langle r \rangle $ and \changemore{(b) the fluctuation}~$\Delta r$  depending on the dynamic field $\Gamma/J_{0}$. The data is shown for $N=13$. }
\end{figure}

One of the key observables in the current study is the variance $(\Delta r)^2=\langle r^2\rangle - \langle r\rangle^2$, where the averages are taken over disorder samples. \changemore{Note that, first, we average the gap ratio over a certain energy window to obtain its value at a fixed disorder realization. Next, we compute the variance of the obtained gap ratios over disorder realizations. The chosen size of the energy window can slightly alter the resulting value of $(\Delta r)^2$, because small window sizes typically introduce additional source of fluctuation. The variance behavior} was previously studied in Ref.~\cite{sierant2019level} in a broader context of the gap ratio distributions between different disorder samples. There, it was observed that the variance has a maximum near the MBL transition point. Hence, it is natural to expect that the maximum fluctuation of the gap ratio emerges at the transition point. Indeed, the transition point is usually estimated as the inflection point in critical parameter ($r$) dependence on the disorder strength (e.g., the ratio $R=\Gamma/J_{0}$).  If we assume that the ratio~$r$ is a function of that parameter, one can represent the fluctuations of $r$ in terms of the fluctuations of $R$ as $\langle\delta r^2\rangle \approx (dr/dR)^2 \langle\delta R^2\rangle$. Since the fluctuation in disorder strength is not strongly sensitive to the transition point, the maximum should be observed at the maximum derivative $dr/dR$ corresponding to the inflection point. 

This expectation is consistent with our studies showing that the maximum fluctuation of the average ratio~$r$ emerges close to the transition region between the chaotic and localized behavior. This is illustrated in Fig.~\ref{fig:Illustration}. 
We suggest using the point of maximum fluctuation $\Delta r$ to determine the phase transition point $\Gamma_{c}(N)$. Note that traditionally the MBL transition point is determined with the finite-size scaling analysis, where the data for $\langle r \rangle$ with different system sizes $N$ are collapsed on one curve with a certain scaling ansatz~\cite{Zhang2021_finite_size_scaling, Wang2021}. This methodology relies on the assumption that for large system sizes, the transition point converges to a certain finite value. Although this may be the case for certain MBL systems with short-range and long-range interactions~\cite{Zhang2021_finite_size_scaling, Wang2021, Lukin2022MBL}, the current analysis of the MBL transition in the Sherrington-Kirkpatrick model points towards the $N$-dependent transition point. Hence, we need a reliable criterion for the transition for the fixed value of $N$. 

From the computational perspective, it is unfeasible to reach large system sizes ($N>14$) using the full exact diagonalization of the Hamiltonian matrix~\eqref{eq:Hrl} partially due to a rather dense structure of the latter (long-range interactions and almost no symmetries) at moderate $N$ and the exponential growth of the matrix size with the further increase of $N$. This brings us to the necessity to employ the filtering approaches, which allow us to diagonalize the Hamiltonian only in the chosen energy window of the full Hilbert space. 

The idea of the filtering follows closely the celebrated Lanczos algorithm to determine the largest (smallest) energy eigenvalues and eigenvectors of the Hamiltonian. The main difference is that instead of the original Hamiltonian $H$, one needs to use a localized function $f(H)$ of the Hamiltonian. This function must be sufficiently small outside of a fixed energy window $E \in [E_{min}, E_{max}]$. There are several possible approaches on how this function can be defined and calculated \cite{Pietracaprina2018, Guan2021a, Guan2021b, Pieper2016, Li2016}. In this study, we closely follow Ref.~\cite{Sierant2020LargeMatr} (see also Ref.~\cite{Sierant2023} for applications to Floquet systems), which proposes the polynomially filtered exact diagonalization method. This method constructs the function $f(H)$ as a finite-order Chebyshev polynomial expansion of the delta function. The number of polynomials in the expansion controls the size of the energy window. Typically, it appears sufficient to choose the energy window covering approximately $500$ eigenstates of the Hamiltonian, precisely in the center of the energy spectrum of the model. Note that the model Hamiltonian has an additional $\mathbb{Z}_{2}$ symmetry upon the simultaneous flip of all spins. This symmetry is used to project the Hamiltonian onto the subspace with an additional eigenvalue of this symmetry generator. This projection reduces the effective Hilbert space dimension twice, and thus improves the speed and memory consumption. 

\begin{figure}[t]
    \centering
    \includegraphics[width=0.6\linewidth]{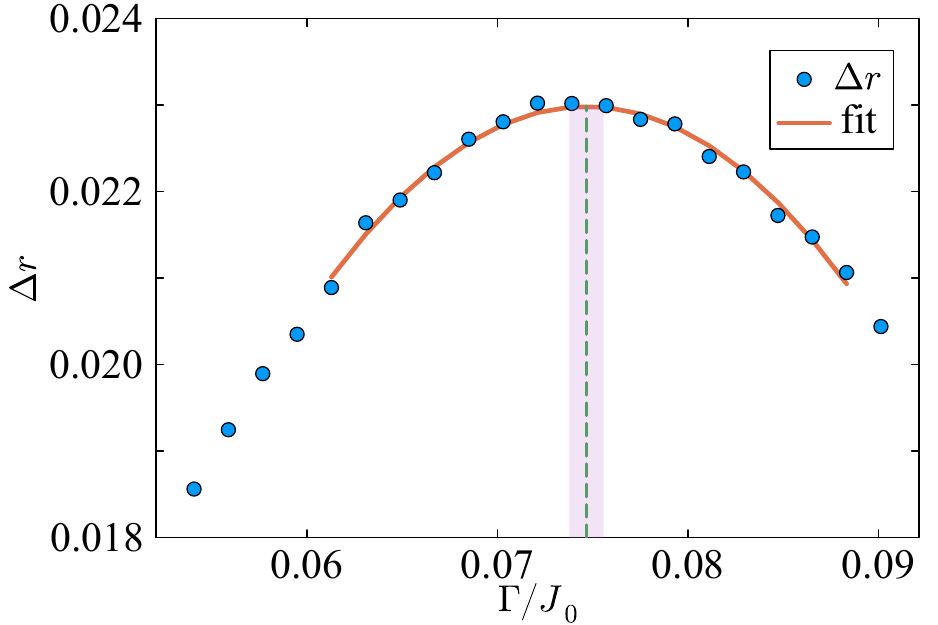}      \caption{\label{fig:Fit}%
        Dependence of $\Delta r$ on $\Gamma/J_{0}$ for $N=13$ and ${\cal N}_{dis}=31200$ (circles) with the corresponding parabolic fit (solid line). The shaded area corresponds to the estimate of the error in determining position of $\Gamma_c$.}
\end{figure}
Our numerical analysis is performed as follows: At the given system size $N \in [12, 19]$, we diagonalize the model~\eqref{eq:Hrl} in a certain energy window for ${\cal N}_{dis}$ samples of the random couplings $J_{ij}$ with the fixed amplitudes $J_0$ and $\Gamma$. We take the number of disorder samples as ${\cal N}_{dis} = \{20000, 31200, 31400, 5400, 800, 400, 400, 400\}$ for $N=\{12,13,14,15,16,17,18,19\}$, respectively. For every choice of couplings, we compute the value of the gap ratio~\eqref{eq:lst} averaged over the predefined energy window. Next, both the mean $\langle r \rangle$ and deviation~$\Delta r$ of this gap ratio are calculated by sampling over different disorder realizations. The resulting quantities are functions of both $N$ and $\Gamma/J_0$. The variance $\Delta r$ as a function of $\Gamma/J_0$ has a clear maximum for every value of $N$. As mentioned above, we take this maximum position as a transition point between chaotic and localized behavior for the given system size $N$. Note that since $\Delta r$ is estimated statistically, it still contains noisy features, which can affect the accuracy of the determination of $\Gamma_{c}$. To minimize the possible influence of these errors, we perform an additional fit of the numerical data $\Delta r(\Gamma/J_0)$ in the vicinity of the maximum with the parabolic function (both data and parabolic fit are illustrated in Fig.~\ref{fig:Fit}). 
To estimate the error bars of each point $\Gamma_c(N)$, we vary the size and position of the window of the fit, and also perform fits with twice smaller numbers of disorder realizations. The spread between the obtained values of the transition point defines the corresponding error bars.  

\begin{figure}[t]
    \centering
    \includegraphics[width=0.6\linewidth]{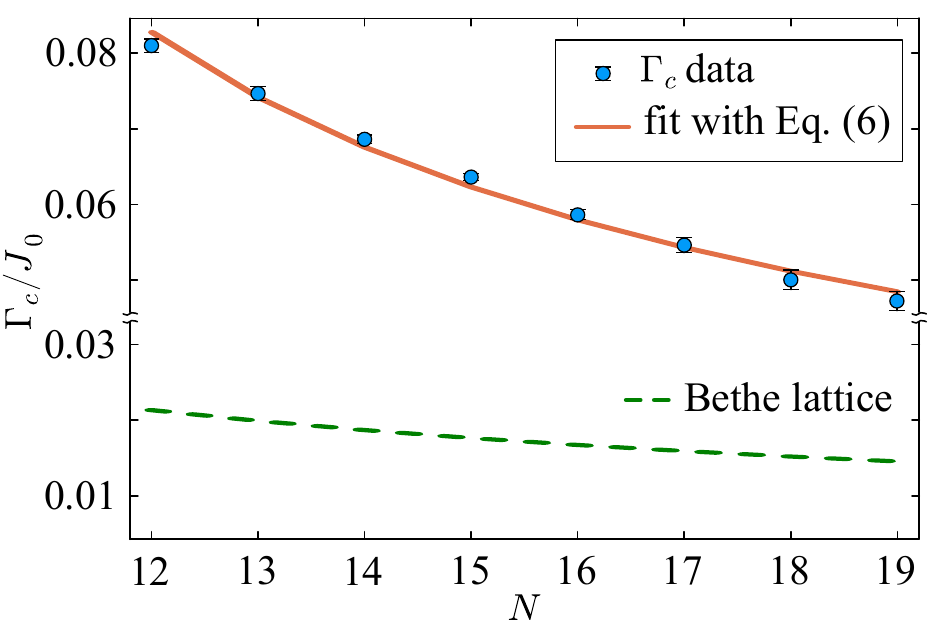}      \caption{\label{fig:Transition}%
        Dependence of the transition point $\Gamma_{c}/J_{0}$ on the system size. The fitting coefficients are $\beta=1.4$ and $\eta=0.41$. }
\end{figure}

Our main goal is to study the dependence of the MBL transition threshold on the system size. We adapt the ansatz in accordance with Eq.~\eqref{eq:MBLThr}, where both $\beta$ and $\eta$ are determined numerically from the least-square fitting of the data points $\Gamma_c(N)$. The best fit predictions are $\beta \approx 1.4$ and $\eta \approx 0.41$. Both numerical results and predictions from the fit are shown in Fig.~\ref{fig:Transition}. In this figure, we also illustrate the results obtained for the Bethe lattice with Eq.~\eqref{eq:BLThr}. It is clear that the localized phase is more stable against the external field $\Gamma$ than in the Bethe lattice model. 
%%%%% FIGURE 4 %%%%% 

\changemore{To compare the present results with those obtained for the Bethe lattice and the quantum random energy model \cite{2025QREM}, we reexpress our criterion using the notations close to that of Refs.~\cite{Parisi_2020,2024RRGKRLSR} as  
\begin{eqnarray}
4\ln\left(\frac{0.16 J_0^2}{\Gamma_c^2}\right)\Gamma_c P(0)N \approx 2.8.
\label{eq:NumThr}
\end{eqnarray}
 Here we replace the argument of logarithm with its squared value having the same dependence on the number of spins $N$ as in the Bethe lattice problem  \changemore{and replace $N$ with $N-1$ since we consider the limit of large number of spins $N$. Particularly, $0.16$ stands for $\eta^2$ and $2.8$ stands for $2\beta$.}    Consequently, the critical quantum field needs to be  approximately $2.8$ times stronger than that for the Bethe lattice to make the states with the zero energy delocalized. We believe that this raise of the localization threshold is due to the destructive interference of contributions of different paths connecting two sites as discussed in Sec.~\ref{sec:analitTh}.  This destructive interference substantially reduces the domain of energies relevant for the  delocalization. The investigation of the quantum random energy model has led to substantially smaller localization threshold~\cite{2025QREM}, possibly  because of the lack of correlations of energies for different paths connecting two sites. 

In the present form, the localization threshold depends on the number spins vanishing within the thermodynamic limit $N \rightarrow \infty$. To make it finite, the interaction of spins can be rescaled similar to that within the Rosenzweig-Porter model \cite{RosenzweigPorterOrig60} (see also Ref. \cite{2025QREM}) as $J_{ij}=\overline{J}_{ij}\sqrt{N}\ln(N)$ ($J_{0}=\overline{J}_{0}\sqrt{N}\ln(N)$). This replacement modifies the definition of localization threshold as $4\Gamma_c \overline{P}(0) \approx 2.8$, where $\overline{P}(0)=2/(\sqrt{2\pi}\overline{J}_{0})$, cf. Eq. (\ref{eq:NumThr}). }

% \change{
% \[
%     P(0) = \frac{2}{\sqrt{2\pi}\sqrt{N-1}J_0} ,
%     \quad
%     P(0)N \approx \frac{2\sqrt{N}}{\sqrt{2\pi}J_0} 
%     =\frac{2\sqrt{N}}{\sqrt{2\pi}\,\overline{J}_{0}\sqrt{N}\ln(N)}
%     = \frac{\overline{P}(0)}{\ln(N)},
% \]
% \[
%     \ln\left(\frac{0.16 J_0^2}{\Gamma_c^2}\right)/\ln(N)
%     =\ln\left(\frac{0.16 N\ln^2(N)\overline{J}_0^2}{\Gamma_c^2}\right)/\ln(N)
%     =1+ 2 \ln\left(\frac{0.4 \ln(N)\overline{J}_0}{\Gamma_c}\right)/\ln(N)\approx 1.
% \]
% }

\begin{figure}[t]
    \centering
    \includegraphics[width=0.6\linewidth]{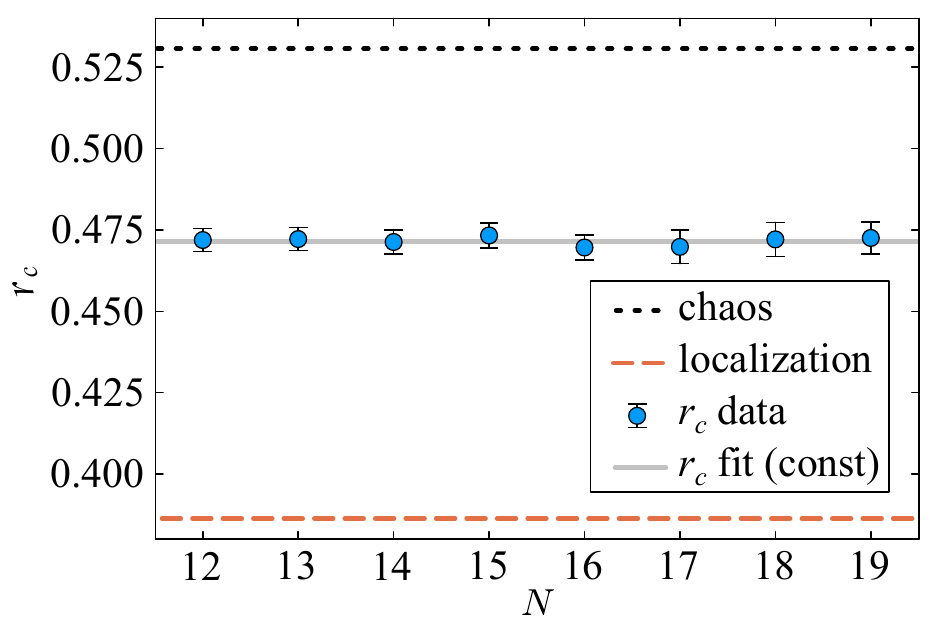}      \caption{\label{fig:rc}%
        Dependence of the value of mean gap ratio at the MBL transition point $r_{c}$ on the system size. The best horizontal fit corresponds to approximately $r_c^{\rm (fit)}\approx0.472$.  }
\end{figure}

\changemore{We} also investigate the correlation between the maximum position of $\Delta r$ and the value of $\langle r \rangle$ at the same critical field $\Gamma_c/J_{0}$, denoted by $r_{c}\approx 0.475$ (see also Fig.~\ref{fig:Illustration}). 
The dependence of $r_{c}$ on $N$ is shown in Fig.~\ref{fig:rc}, where it exhibits nearly constant behavior in the studied range of system sizes. \changemore{This is similar to the behavior of the critical ratio for the quantum random energy model \cite{2025QREM} at an infinite temperature, though that parameter is closer there to the localization side ($r_c\approx 0.415$). For random regular graphs organized similarly to the Bethe lattice, the critical gap ratio $r_c$ approaches its Poisson limit corresponding to the localized states \cite{sierant2023RRG}. }
%%%%% FIGURE 5 %%%%% 

\changemore{To provide an alternative description of the localization transition, we can also represent the transition points at $r=r_c$ as the crossing points of the $r$ curves. Following  Ref.~\cite{ab16SG}, we express the transverse field in terms of the rescaled amplitude as $(\Gamma-\Gamma_c)/\delta \Gamma_c$, where we take $\Gamma_{c}$ from the analysis of variance $\Delta r$. Next, we assume the universal scaling $\delta \Gamma_c =\Gamma_c / N^{\nu}$, and seek for the optimal exponent $\nu$. We define the loss function as the accuracy of fitting of the rescaled curves with a single polynomial function of order 4. The corresponding least-square numerical analysis of this loss function with the step size $\Delta \nu=0.01$ points to $\nu=1.49$, which is very close to the rational number $3/2$. With this optimal exponent, the graphs for different system sizes collapse reasonably well to a single universal curve shown in Fig.~\ref{fig:6}(a). To confirm this finding, in Figs.~\ref{fig:6}(b) and \ref{fig:6}(c) we also show the dependencies of the gap ratio on the transverse field with $\nu=1$ and $\nu=2$, respectively.
%%%%% FIGURE 6 %%%%% 
\begin{figure}[t]
    \centering
    \includegraphics[width=\linewidth]{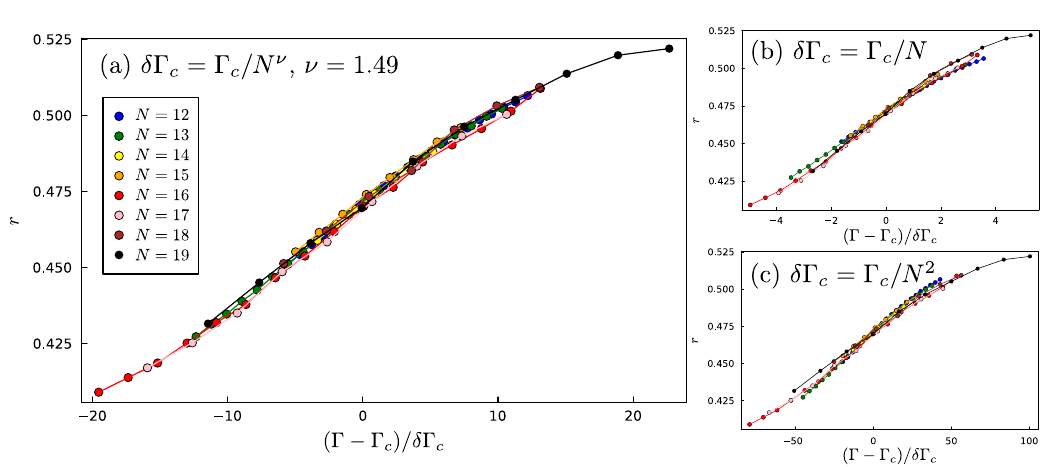}\caption{\label{fig:6}%
        \changemore{Dependence of the gap ratio on the field $\Gamma$ with different scaling behavior of the critical domain size~$\delta\Gamma_c$ on $N$.}
        }
\end{figure}
The collapse of the $r$ dependencies at $\nu\approx1.5$ supports the relevance of our definition of localization threshold based on the maximal fluctuation of the gap ratio~$r$. This definition can be hopefully used for other problems of interest. 

The estimated exponent differs from the previously estimated exponents $\nu \approx 1$ both for the $XY$ model with the long-range interactions \cite{ab19Quartets}, quantum random energy model \cite{2025QREM} and random regular graphs \cite{sierant2023RRG}.  The difference could be originated from the larger number of random parameters, i.e., $N(N-1)/2$ random interactions in the present work instead of $N$ random potentials in Refs. \cite{ab19Quartets,2025QREM}. Averaging over larger numbers of parameters can reduce the fluctuations.    
} 

\section{Conclusion}
\label{sec:Concl}

\changemore{We have investigated the many-body localization (MBL) transition in the quantum Ising model with long-range, distance-independent interactions using both analytical arguments and numerical simulations. In contrast to the MBL problem with short-range interactions, where strong finite-size effects hinder reliable estimates,  we find that the numerically extracted localization threshold is consistent with the analytical predictions.

Our results are compared with those for the matching Bethe-lattice problem and for the quantum random energy model. We find that the critical dynamical field $\Gamma_{c}$ for the MBL transition in our system exceeds the corresponding thresholds in both models by a factor of order three. This enhancement arises from the destructive interference of spin-flip processes, which is absent in the other two models. Thus, the present problem occupies an intermediate position between single-particle localization models (for example, the quantum random energy model on the Bethe lattice, which is equivalent to Anderson localization on the hypercube) and paradigmatic many-body localization models with short-range interactions. We hope that the results reported here, together with future investigations, will shed further light on the nature of the MBL transition. Our approach may also be extendable to physically relevant systems with long-range interactions decaying as a power law with distance.} 

The analytical expression for the transition, Eq.~(\ref{eq:MBLThr}), accurately describes the dependence of the critical field on system size, as confirmed by exact diagonalization when parameters $\beta = 1.4$ and $\eta = 0.41$ are used. Furthermore, the weak size dependence of the average ratio $r$ at the transition (Fig.~\ref{fig:rc}) indicates that finite-size effects are already minimal for $N=18$ spins, yielding reliable estimates of the transition point. This contrasts sharply with MBL in systems with short-range interactions~\cite{Sierant2022Challenges}, where much larger system sizes are required. Thus, both analytical and numerical results demonstrate that the MBL transition in the quantum Ising model with infinite-range interactions can be characterized reliably using the Bethe lattice framework, providing a powerful model probe of critical behavior near the transition.

\changemore{For each number of spins $N$, the localization threshold is estimated from the maximum fluctuation of the gap ratio $r$. The resulting thresholds are consistent with the expected collapse of the rescaled $r(\Gamma)$ curves, confirming the reliability of this approach. Consequently, these relatively simple computational methods can be applied to other systems of interest to probe their localization thresholds.

In our model of size-independent spin-spin interactions of the order of $J_{0}$, we found that the critical transverse quantum field scales with the number of spins as $\Gamma_{c}\propto J_0/(\sqrt{N}\ln(N))$, while for the normalized interaction $J_{0}/\sqrt{N}$ \cite{Sherrington75} it scales as $\Gamma_{c}\propto J_0/(N\ln(N))$. The width of the critical domain scales with the number of spins as $\delta \Gamma_{\rm c}/\Gamma_{\rm c} \sim N^{-1.5}$. }

\section*{Acknowledgements}

We are  grateful to Denys Bondar for useful discussions and help with using OSF and Jan Kune\v{s} for assistance with the execution of the numerical part of the project.  

% TODO: include author contributions
%\paragraph{Author contributions}
%This is optional. If desired, contributions should be succinctly described in a single short paragraph, using author initials.

% TODO: include funding information
\paragraph{Funding information}
This work is supported by the Tulane University Lavin Bernick Fund and by the Ministry of Education, Youth and Sports of the Czech Republic through the e-INFRA CZ (ID:90254), project OPEN-34-73.
I.L. acknowledges support by the IMPRESS-U grant from the US National Academy of Sciences via STCU project No.~7120 and  the IEEE program ‘Magnetism for Ukraine 2025’, Grant No.~9918.
A.S. acknowledges support by the National Research Foundation of Ukraine, project No.~0124U004372. A.B. also acknowledges support by the National Science Foundation (CHE-1462075).  

%Authors are required to provide funding information, including relevant agencies and grant numbers with linked author's initials. Correctly-provided data will be linked to funders listed in the \href{https://www.crossref.org/services/funder-registry/}{\sf Fundref registry}.

\begin{appendix}
%\numberwithin{equation}{section}
\changemore{
\section{Estimate of the localization threshold for the Bethe lattice with the Gaussian disorder}
\label{App:BLThr}

Here we derive the approximate analytical expression for the localization threshold for the Bethe lattice problem with the Gaussian distribution of site energies characterized by the width~$W$ and large coordination number $N$. In the limit of a large coordination number $N$, a real part of self-energy can be approximately neglected near the localization threshold  \cite{AbouChacra73}. Next, we use the Gaussian distribution of random energies $P(x)$  replacing  the function $Q$ in \cite[Eq.~7.7]{AbouChacra73} as 
\begin{eqnarray}
    1 &\approx& -2(N-1)\Gamma_{\rm B}\int_{0}^{\infty}(P'(x)-P'(-x))\ln\left(\frac{x}{\Gamma_{\rm B}}\right)dx,  
    \label{eq:AppBLThr0}
\end{eqnarray}
where $P'(x) =dP/dx$. Using the identity $P'(x)=-P'(-x)$ and representing the logarithmic term in the form $\ln(x/\Gamma_{\rm B})=\ln(x/W) +\ln(W/\Gamma_{\rm B})$, we evaluate the integral containing the first term numerically as $0.0462/W$ and the second integral analytically as $2\ln(W/\Gamma_{\rm B})/(\sqrt{2\pi}W)$. Combining these terms together we end up with the approximate definition of the localization threshold in the form used in the main text,
\begin{eqnarray}
    1 &\approx& 4(N-1)\frac{\Gamma_{\rm B}}{\sqrt{2\pi}W}\ln\left(\frac{1.06W}{\Gamma_{\rm B}}\right). 
    \label{eq:AppBLThr}
\end{eqnarray}
}
%Add material which is better left outside the main text in a series of Appendices labeled by capital letters.

%\section{About references}
%Your references should start with the comma-separated author list (initials + last name), the publication title in italics, the journal reference with volume in bold, start page number, publication year in parenthesis, completed by the DOI link (linking must be implemented before publication). If using BiBTeX, please use the style files provided  on \url{https://scipost.org/submissions/author_guidelines}. If you are using our LaTeX template, simply add
%\begin{verbatim}
%\bibliography{MBL}
%\end{verbatim}
%at the end of your document. If you are not using our LaTeX template, please still use our bibstyle as
%\begin{verbatim}
%\bibliographystyle{SciPost_bibstyle}
%\end{verbatim}
%in order to simplify the production of your paper.
\end{appendix}


% Provide your bibliography here. You have two options:

%%% FIRST OPTION
% Write your entries here directly, following the example below, including:
% Author(s), Title, Journal Ref. with year in parentheses at the end, followed by the DOI number.

%\begin{thebibliography}{99}
%\bibitem{1931_Bethe_ZP_71} H. A. Bethe, {\it Zur Theorie der Metalle. i. Eigenwerte und Eigenfunktionen der linearen Atomkette}, Zeit. f{\"u}r Phys. {\bf 71}, 205 (1931), \doi{10.1007\%2FBF01341708}.
%\bibitem{arXiv:1108.2700} P. Ginsparg, {\it It was twenty years ago today... }, \url{http://arxiv.org/abs/1108.2700}.
%\end{thebibliography}

% Use your bibtex library, formatted by the SciPost style file.
%\bibliography{SciPost_Example_BiBTeX_File.bib}
\bibliography{MBL.bib}

%%%%%%%%%% END TODO: BIBLIOGRAPHY

\end{document}